\documentclass[aps,prb,twocolumn,groupedaddress,showpacs]{revtex4}
\usepackage{amssymb,amsmath,graphicx}

\newcommand{\mx}{\text{m}}

\begin{document}



\title{Single-color pyrometry of individual incandescent multiwalled carbon nanotubes}


\author{S. B. Singer}

\author{Matthew Mecklenburg}

\author{E. R. White}

\author{B. C. Regan}
\email{regan@physics.ucla.edu}
\affiliation{%
Department of Physics and Astronomy, University of California, Los Angeles, California, 90095, USA\\
and California NanoSystems Institute, Los Angeles, California, 90095, USA
}


\date{\today}

\begin{abstract}
Objects that are small compared to their thermal photon wavelengths violate the assumptions underlying optical pyrometry and can show unusual coherence effects.  To investigate this regime we measure the absolute light intensity from individual, incandescent multiwalled carbon nanotubes.   The nanotube filaments' physical dimensions and composition are determined using transmission electron microscopy and their emissivities are calculated in terms of bulk conductivities.  A single-color pyrometric analysis then returns a temperature value for each wavelength, polarization, and applied bias measured.  Compared to the more common multiwavelength analysis, single-color pyrometry supports a more consistent and complete picture of the carbon nanotube lamps, one that describes their emissivity, optical conductivity, and thermal conductivity in the range 1600--2400~K.
\end{abstract}



\maketitle

\section{Introduction}
Above the melting point of silver (1234.93~K) the International Temperature Scale of 1990 (ITS-90) defines the best practical estimates of thermodynamic temperature with single-color optical pyrometry.\cite{1990ITS}  Pyrometry is chosen for implementing the ITS-90 at high temperatures because of its relative ease of application and its well-understood connection to temperature via Planck's law. In fact, because Planck's law is material-independent and valid at any temperature, in principle pyrometry is a universal thermometric technique. However, Planck's derivation explicitly assumes that the radiating object's dimensions are large in comparison to a typical thermal wavelength.\cite{1959Planck}  Objects in the opposite limit violate this assumption, which leads to coherence effects that make the connection between temperature and the emitted radiation much less straightforward.\cite{2005ChenGangBook,2008Carey}  Sub-wavelength structure in thermal radiators has recently been shown to lead to such unfamiliar effects as directed  radiation \cite{2002Greffet} and enhanced coherence lengths.\cite{2008Klein} A large object described by the Planckian picture radiates incoherently from its well-defined surface area in the lowest approximation, but a small object contributes coherently with its whole volume to the electromagnetic field. Thus a sufficiently small object radiates power in proportion to its volume, not its surface area, and can have an emission efficiency that exceeds the blackbody ``limit''.\cite{1983Bohren}  By altering these scaling laws and hence the absolute magnitude of the radiation emitted, size effects directly impact how thermodynamic temperature is defined and determined for small objects.

Here we report the application of single-color (or monochromatic) optical pyrometry to individual incandescent carbon nanotubes.  In addition to being small, carbon nanotubes (CNTs) are stable at high temperatures \cite{2001Collins,2002Miyamoto,2002Purcell,2002Sveningsson,2003LiPol,2004Cai,2005Sveningsson,2006Wei,2007Begtrup1,2007ZhangCO2,2007ZhangMelting,2008Aliev,2009Deshpande,2010Lim,2011Akita,2011Liu,2011Natarajan} and thus are of interest for many applications, including field emission tips,\cite{2002Purcell,2005Sveningsson,2006Wei} nanoscale heaters,\cite{2011Akita} and incandescent lights.\cite{2003LiPol,2009Liu} The temperatures $T$ that CNTs reach have been inferred from the magnitude of their field emission current,\cite{2002Purcell} their breakdown temperature,\cite{2005Chiu}, the melting of nearby gold nanoparticles,\cite{2007Begtrup1} $T$-dependent Raman spectroscopy,\cite{2008Hsu,2009Deshpande} scanning thermal microscopy,\cite{2000Shi} microfabricated Pt thermometers,\cite{2001Kim} and their current-voltage response function.\cite{2007Mann}   However, multiwavelength optical pyrometry is the most common thermometry technique applied to CNTs at high temperatures.  This general, non-contact method allows the detector to be remote from the source and is relatively easy to implement above $1000$~K.  Thus it has been applied to individual CNTs, \cite{2004Cai,2011Liu}  films,\cite{2002Sveningsson,2007ZhangCO2,2008Aliev,2009Liu,2010Lim,2011Natarajan} and bundles.\cite{2003LiPol,2006Wei}

But while some variant of optical pyrometry is often preferred for measuring high temperatures, in practice the method is challenging to implement accurately because of uncertainties associated with the radiating object's emissivity.\cite{1981Coates,1985Hunter,2001Neuer} The multiwavelength pyrometry measurements described in Refs.~(\onlinecite{2002Sveningsson,2007ZhangCO2,2008Aliev,2009Liu,2010Lim,2011Natarajan,2003LiPol,2006Wei,2004Cai,2011Liu}) assume that the nanotube source is a greybody, $i.e.$ that its emissivity $\varepsilon$ is independent of wavelength.  The greybody assumption simplifies the thermometry enormously, for then neither detailed knowledge of the source geometry nor an absolute calibration of the light collection apparatus is required; only the relative intensities matter.  However, spectral and polarization-dependent features have been observed in thermal spectra of both CNT films\cite{2007ZhangCO2, 2008Aliev,2010Lim} and individual CNTs\cite{2004Cai,2011Liu,2011SingerPol}. Thus CNTs are known to have non-trivial emissivities that will affect multiwavelength pyrometric measurements.

Previously we have calculated the polarization-dependent emission efficiency of multiwalled carbon nanotubes (MWCNTs).\cite{2011SingerPol} The net polarization depends only on the ratio of the efficiency parallel and perpendicular to the nanotube axis.  Here we use the absolute magnitude of these efficiencies, as opposed to the ratio, to determine the temperature of individual incandescent MWCNTs as a function of input electrical power.  Then single-color pyrometric measurements, which are modeled on the recommendations for approximating the ITS-90, give a separate temperature determination for each wavelength and polarization detected.  Comparing these results  allows us to quantitatively evaluate the accuracy of our temperature determination.  We find that a self-consistent picture emerges where a MWCNT's maximum temperature is weakly exponential in the applied electrical power, its normal emission efficiency is $\sim 1$ for the parallel polarization and $\sim 0.1$ for the perpendicular polarization, its optical conductivity is approximately that of graphene, and its thermal conductivity decreases by a factor of $\sim 2$ between room temperature and 2000~K.

\section{Experiment and Theory}
\begin{figure}
\begin{center}
\includegraphics[width=.8 \columnwidth]{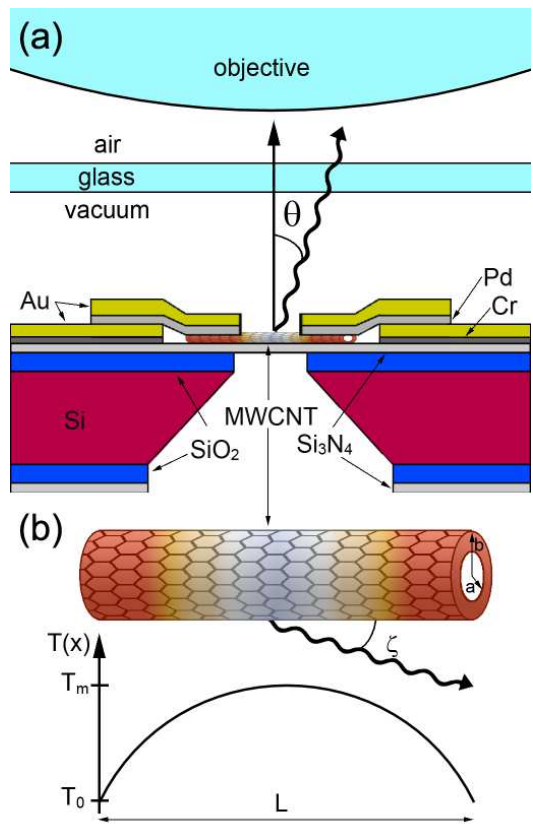}
	\caption{\label{fig:cartoon} (Color online) (a) Nanolamp device architecture and optical measurement configuration (not to scale). The MWCNT and an electron-transparent silicon nitride membrane span a hole in a silicon chip.  Applying a bias voltage to the lithographically-defined metal contacts heats the nanotube to incandescence.  The light is collected by a microscope objective positioned outside the high vacuum chamber. (b) Schematic showing the geometric variables defining the Mie model of the nanotube, and a  plot indicating the parabolic temperature distribution supported by the nanotube under bias.}
\end{center}
\end{figure}

The experimental procedure, including the arrangement of the light collection apparatus and the fabrication of the MWCNT devices, has been described previously\cite{2009Fan,2011SingerPol}, but we give a brief summary here (also see Fig~\ref{fig:cartoon}a).  Individual arc-discharge grown MWCNTs suspended on electron-transparent silicon nitride membranes are contacted via e-beam lithography. After its dimensions have been determined in a transmission electron microscope (TEM), a MWCNT is brought to incandescence in vacuum by applying a voltage bias.  A $100\times$ microscope objective (numerical aperture NA$=0.5$) collects the light and forms a diffraction-limited image (resolution 0.42$\lambda$) on a cooled CCD camera capable of single-photon detection (Princeton Instruments Pixis 1024BR).  With the aid of fourteen 10-nm bandpass color filters and a Wollaston prism, thermal light emission is measured as a function of applied bias, wavelength, and polarization.

As thermal light sources individual MWCNTs lie in the unusual, intermediate size regime of interest. MWCNTs are large compared to atoms, small molecules, and even single-walled CNTs, in that they have many degrees of freedom and are not expected to show strong spectroscopic features.  However, a typical MWCNT is small compared to a typical thermal photon wavelength $\lambda^\gamma_T\sim \hbar c/(k_B T)$ in at least two of its three dimensions.  (The nanotubes reported here have outer diameters $\sim 15$~nm, lengths  $\sim 1$~$\mu$m, and $\lambda^\gamma_T \gtrsim 1$--3~$\mu$m.)  Because MWCNTs are thermodynamically ``large'' it is reasonable to treat them as having bulk electrical and thermal conductivities, dielectric constants, \emph{etc.} in the lowest order approximation. Furthermore, a typical thermal phonon wavelength $\lambda^p_T\sim \hbar v_\text{sound}/(k_B T)$ is comparable to the interatomic spacing, which means that a local temperature can be defined. However, because MWCNTs are electrodynamically ``small'' they do not have a well-defined radiating surface area, but rather contribute with their whole volume to the thermal radiation field.

To provide a bridge between the ``large'' object, geometric optics picture and the ``small'' object, physical optics picture,\cite{1978Wolf} it is useful  \cite{1983Bohren} to define an emission (equivalent to absorption by Kirchhoff's law) cross section $C(\lambda,p,\Omega)$, which is a function of wavelength $\lambda$, polarization $p$, and emission direction $\Omega=(\theta, \phi)$.  This cross section plays the role normally filled by the product $A \varepsilon (\lambda,p,\Omega)$ where $A$ is the surface area and $\varepsilon$ is the emissivity. Thus, for emitters with dimensions small compared to a wavelength, the number of thermal photons with polarization $p$, per bandwidth $d \lambda$, per unit solid angle $d \Omega$, emitted in direction $\Omega$ is given by
\begin{equation}\label{eq:geoopt}
\frac{\dot{N}}{d\lambda d\Omega}  = C(\lambda,p,\Omega) \frac{c}{\lambda^4} \frac{1}{e^{C_2/\lambda T} - 1},
\end{equation}
where $c$ is the speed of light in vacuum and $C_2 = h c/k_B=14.4$~$\mu\text{m}\cdot\text{kK}$ is the second radiation constant.  For ease of comparison with the geometric optics limit,  the cross section $C$ is split into a product $C=A' Q$, where $A'$ is the ``hard'' projected area in the normal direction (as determined by TEM, for instance) and $Q$ is the efficiency.  The efficiency is the small-radiator equivalent of the emissivity, but unlike the emissivity its values are not constrained to be less than unity. \cite{1983Bohren}  The efficiency also includes the geometric factors that account for the relative orientation of the emitting object and emission direction, \emph{e.g.} a cosine in the case of a Lambertian emitter. 

To extract the nanotube temperature from a measurement of the emitted light intensity, we require a detailed model of the relationship between the nanotube temperature distribution and the image captured by the CCD.  The calculation proceeds in two steps. First, we map the nanotube's three-dimensional radiation field onto a fictitious two-dimensional object plane, including efficiency effects from the object, the optics, and the camera.  Second, we map the object plane intensity onto the CCD in the image plane by convoluting it with the microscope point spread function (PSF), which we approximate with a Gaussian.\cite{2007ZhangPSF}  

As a function of position, polarization, and filter, the object plane intensity $\dot{O}_{p,f}(x_o,y_o)$ for a nanotube of outer radius $b$ oriented along the $x$-axis, in units of counts per second per area, is 
\begin{align}\label{eq:Object}
\dot{O}_{p,f}(x_o,y_o)&= H(b+y_o) H(b-y_o)\nonumber\\
& \times 
\frac{c}{\lambda^4} \, \frac{\Theta_f \Delta \lambda}{e^{C_2/\lambda T(x_o)} - 1} \,\int_{\text{NA}} Q(\Omega,\lambda,p) d\Omega,
\end{align}
where $H$ is the Heaviside step function, $\Theta_f$ is the quantum efficiency of the optical system for filter $f$, and  $\Delta \lambda$ is the filter bandwidth.  The function $T(x_o)=T_{\mx}+(T_0 - T_{\mx})(2 x_o/L)^2$ in the Planck factor describes the parabolic temperature distribution supported by a MWCNT device of length $L$ under bias;\cite{2009Fan} as shown in Fig.~\ref{fig:cartoon}b, the  temperature peaks at the midpoint ($T_{\mx}$) and decreases to $T_0$ at the contacts.

In the convolution integral we approximate the product $H(y_o-b) H(y_o+b)$ as $2b \delta(y_o)$ since the nanotube outer radius $b$ is small compared to the effective CCD pixel size $\beta$ and the PSF width $s=0.21 \lambda/\text{NA}$.  Integrated over a CCD pixel labeled by integers $(i,j)$, the PSF has the Gaussian form 
\begin{equation}\label{eq:PSF}
\text{PSF}(i,j,x_o,y_o)=\beta^2\frac{\text{NA}^2 \pi}{\lambda^2} e^{\frac{(i\beta-x_o)^2 +(j \beta-y_o)^2}{-2 s^2}},
\end{equation}
where we have taken the mean intensity over a pixel to be equal to the central intensity. The normalization chosen for Eq.~(\ref{eq:PSF}) gives an excellent approximation (within $6\%$) of the actual Airy PSF within $0.75\lambda$ of $(x_o,y_o)$, although it underestimates the total signal in the image plane by a factor $2 \pi^2 (0.21)^2\simeq 0.87$.

We also allow for an arbitrary position and orientation of the nanotube relative to the CCD coordinate system, writing the nanotube coordinates $(x_o,y_o)=(\ell \cos\phi+x_i,\ell\sin\phi+y_i)$, where $(x_i, y_i)$ describe the nanotube midpoint's displacement from a pixel center, $\ell$ varies in $[-L/2,L/2]$, and $\phi$ is the angle the nanotube makes with respect to the CCD axes. Convoluting Eq.~(\ref{eq:Object}) with Eq.~(\ref{eq:PSF}) then gives the CCD count rate, \cite{2009Fan} 
\begin{equation}\label{eq:counts}
\begin{split}
\dot{S}(i,j) \simeq 0.21\sqrt{8 \pi^3} \text{NA}\,& Q_{\text{NA}}\, \tilde{\text{PSF}}_{ij} \Theta_f \\
&\times \frac{b}{\lambda} \frac{\Delta \lambda}{\lambda} \frac{c \beta^2}{\lambda^3} \eta e^{-C_2/\lambda T_\mx},
\end{split}
\end{equation}
where the image eccentricity $\eta$ is given by
\begin{equation}
 \eta \equiv 1/\sqrt{1+ 8 C_2 (T_\mx-T_0) s^2/(\lambda T_\mx^2 L^2)},
\end{equation}
and $Q_{\text{NA}}= \int_{\text{NA}} Q d\Omega$ is the integral of the efficiency over the numerical aperture.  The factor $\tilde{\text{PSF}}_{ij}$ describing the variation in count rate with position on the CCD is
\begin{equation}\label{eq:PSFtilde}
\begin{split}
\tilde{\text{PSF}}_{ij} = \exp \{ &-[ (i\beta-x_i)^2 (1 - \eta^2 \cos^2 \phi) \\
&+ (j\beta-y_i)^2 (1 - \eta^2 \sin^2 \phi) \\
&- 2 \eta^2 (i\beta - x_i) (j \beta - y_i) \sin \phi \cos \phi ]/2s^2\}.
\end{split}
\end{equation}
A single CCD exposure gives $\dot{S}(i,j)$, which we invert using Eq.~(\ref{eq:counts}) to find the peak temperature $T_\mx$.

To solve for $T_\mx$ all of the other variables must be determined.  The central wavelength $\lambda$ and the bandpass $\Delta \lambda$ are specified by the optical filter supplier (Chroma Technology).  For all filters the nominal bandpass is 10~nm, within the range recommended for realizing the ITS-90. The effective pixel size $\beta=127$~nm is the physical pixel size of 13~$\mu$m divided by the system magnification, which we find to be 2\% bigger than its nominal value of 100 by imaging a Ronchi ruling.  The PSF width $s=0.42 \lambda$ is verified by fitting the image spatial intensity to a 2D Gaussian function.\cite{2009Fan}  The remaining parameters are $\Theta_f$, the nanotube geometry variables, $Q_{\text{NA}}$, and $\eta$. 

We determine the collection efficiency $\Theta_f$ in counts per photon by imaging a 100~$\mu$m-diameter stainless steel pinhole (Thorlabs) illuminated from behind by a 45~W calibrated tungsten lamp at a distance of 50~cm.  The pinhole diameter is chosen to be smaller than the microscope field of view to allow focusing, but larger than the wavelengths of interest and the pinhole thickness (12.5~$\mu$m) to minimize diffraction effects.  The lamp supplier (Newport Corporation) specifies the spectral irradiance $\Lambda$ in mW/m$^2$~nm at this distance with an uncertainty of 3\%.  Measurements of the collected signal as a function of exposure time under constant illumination conditions verify that the non-linearity of the CCD is less than 1\%, as specified by the camera manufacturer.  Averaging over $\sim 10^5$ illuminated pixels, we find $\dot{S}_{45\text{W}}$, which is related to $\Lambda$ and $\Theta_f$ by 
\begin{equation}
\dot{S}_{45\text{W}} =\beta^2  \frac{ \lambda \Lambda(\lambda)}{h c}\,\Delta \lambda\, \Theta_f.
\end{equation}
Since the lamp subtends a small solid angle at this distance from the objective, the measurement of $\dot{S}_{45\text{W}}$ gives a value for the product $\Delta \lambda\, \Theta_f$ that is strictly valid only near normal incidence.  Light rays exit the objective at small angles relative to the optical axis, so this distinction is negligible except at the vacuum window and objective lens.  However, the Fresnel equations indicate that the error introduced even for rays at the edge of the numerical aperture is always $\lesssim 10~\%$.

\begin{figure}
\begin{center}
\includegraphics[width= \columnwidth]{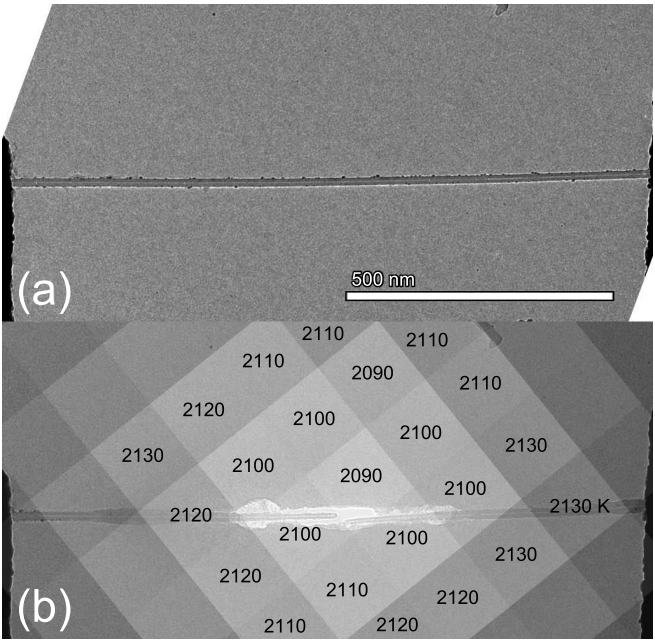}
	\caption{\label{fig:map} (a) TEM image of a nanolamp device prior to biasing.  The nanotube is 1.16~$\mu$m long and has 17 walls. (b)  Superposition of a post-illumination TEM image and optical data from the same device collected at maximum bias in the parallel polarization with the $\lambda= 500$~nm color filter. Due to the high temperatures achieved the membrane near the nanotube midpoint has failed, and the nanotube is broken. The squares correspond to pixels on the CCD (effective dimension $\beta=127$~nm).  Pixels near the nanotube midpoint are labeled with $T_\mx$ values in kelvins as calculated by inverting Eq.~(\ref{eq:counts}).  Note that these numbers refer not to the temperatures at their different locations, but rather to the temperature at the nanotube midpoint only.}
\end{center}
\end{figure}

The nanotube outer radius $b$ and length $L$ (see Fig.~\ref{fig:cartoon}b) are determined with an FEI Titan 80-300 TEM operating at an accelerating voltage of 80~kV. The images captured confirm the composition and structure of the MWCNT radiator, \emph{i.e.} concentric, tubular sheets of pure carbon.  The TEM also determines the inner radius $a$ and the number of walls $n$, variables that figure into the calculation of $Q$ and cannot be determined with other microscopy techniques, \emph{e.g.} SEM, AFM, or STM. The parameters $x_i$, $y_i$, and $\phi$ describing the nanotube position relative to the CCD coordinate system are determined by aligning a sequence of optical and TEM images of the device.\cite{2009Fan}    Figure~\ref{fig:map}a shows a TEM image of a representative nanolamp.  
	
The filament's efficiency $Q^p(\theta=\pi/2-\zeta,\phi)$ we have calculated previously \cite{2011SingerPol} using classical Mie theory, modeling the MWCNT as a hollow right cylinder with a complex dielectric constant $\epsilon$. Because the nanotube and optical axes are orthogonal (see Fig.~\ref{fig:cartoon}), we perform the angular integral  $\int_{\text{NA}} Q d\Omega $ numerically.  To illustrate how the efficiencies depend upon the filament conductivity, we here give their normal-incidence values $Q(0,\phi)\equiv Q(0)$ with Re$(\epsilon)=1$ and Im$(\epsilon)$ set by the optical conductance of graphene, $\sigma_g = \pi \alpha/Z_0$ (Refs.~\onlinecite{2002Ando,2006GusyninPRB,2008Mak,2008Nair,2010Tree}):
\begin{equation}\label{eq:efficiency}
\begin{split}
 Q^{||}(0) &\approx n \pi^2 \alpha \approx .072 n,\\
 Q^{\perp}(0) &\approx \frac{16 \pi^2 n \delta^2}{\alpha \lambda^2} \approx .005 n \left(\frac{700~\text{nm}}{\lambda}\right)^2,
\end{split}
\end{equation}
where $\alpha \simeq 1/137$ is the fine structure constant, $Z_0 \simeq 377~\Omega$ is the impedance of free space, $\delta \simeq 0.34$~nm  is the interwall spacing, and $n$ is the number of walls.  Both efficiencies $Q$ are proportional to $n$, which implies that the cross sections are proportional the tube volume, as expected for a ``small'' object.  The nanotubes reported in this study have dimensions and wall numbers such that  $Q^{||}(0) \sim 1$ and $Q^{\perp}(0) \sim 0.1$.  For $n \geq 14$ walls, $Q^{||}(0) > 1$, giving a cross section that is larger than the geometric cross section.  Note also that, as the filament conductivity $\propto \alpha$ increases, $Q^{||}(0)$ increases while $Q^{\perp}(0)$ decreases.  This stark difference in qualitative behavior provides a strong consistency check, as we will show later.

With all of the variables determined Eq.~(\ref{eq:counts}) has no free parameters. A fit is not necessary and $T_\mx$ can be calculated directly (compare Refs.~\onlinecite{2002Sveningsson,2003LiPol,2007ZhangCO2,2010Lim,2011Natarajan,2011Liu,2008Ward}). In the limit that the tube length $L$ is large the eccentricity $\eta \rightarrow 1$ and the count rate Eq.~(\ref{eq:counts}) reduces to the simple form
\begin{equation}\label{eq:simple}
\frac{\dot{S}(i,0)}{\dot{S}_{45\text{W}}}\simeq\frac{h c^2}{\lambda^5 \Lambda(\lambda)}\frac{0.21 b \,\text{NA}\,\sqrt{8\pi^3}}{\lambda}  Q_{\text{NA}} e^{-C_2/\lambda T_\mx} ,
\end{equation}
on axis for a centered, CCD-aligned nanotube, which shows that the temperature determination is not sensitive to the precise value of the filter bandwidth $\Delta \lambda$ or the effective pixel size $\beta$.  Our devices are not in the large $L$ limit;  the implicit dependence of $\eta$ on $T_\mx$ makes Eq.~(\ref{eq:counts}) a transcendental equation which we solve numerically for 16--64 pixels (a square of dimension $\sim \lambda$, corresponding to a count rate reduction by a factor $\lesssim 6$) around the brightest pixel. Figure~\ref{fig:map}b shows how different pixels in a representative optical image give consistent values for $T_\mx$, and how the image plane (CCD pixel) signal intensities $\dot{S}(i,j)$ correspond to object plane (nanolamp device) locations. 

\section{Results}
The inversion of Eq.~(\ref{eq:counts}) gives values of $T_\mx(P,i,j,\lambda,p)$ as a function of input electrical power $P$, pixel location $(i,j)$, wavelength $\lambda$, and polarization $p$.  If the nanotube is in a steady state with a fixed value of $P$,  it is reasonable to combine measurements corresponding to different $(i,j)$, $\lambda$, and $p$ to arrive at a best estimate $\bar{T}_\mx$ for the temperature at the midpoint.  

To find $\bar{T}_\mx(P,\lambda,p)$ from the data in a single image we weight the $T_\mx$ value from each pixel according to the shot noise error $\propto \sqrt{S(i,j)}$.  Based on the number of photoelectrons captured per CCD pixel $N_e\gtrsim 10^4$ we expect signal fluctuations $1/\sqrt{N_e}\lesssim 1\%$, which according to Eq.~(\ref{eq:simple}) implies a relative error $\delta T_\mx/T_\mx\sim (\lambda T_\mx/C_2)/\sqrt{N_e}$.  The observed deviations are $\delta T_\mx/T_\mx\sim 1\%$, which may reflect the approximations inherent in our model.  Nonetheless this spatial uniformity  of the temperature determination shows that the model summarized by Eqs.~(\ref{eq:counts}--\ref{eq:PSFtilde}) is reasonable at the percent level.
 
\begin{figure}
\begin{center}
\includegraphics[width= \columnwidth]{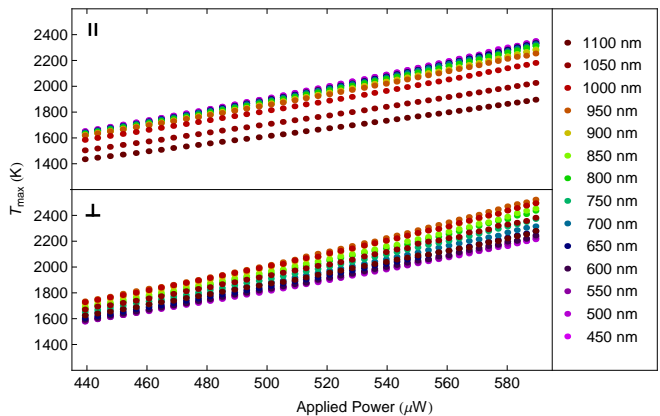}
\caption{\label{fig:TvsP} (Color online) The midpoint temperature $T_\mx$, as determined for each wavelength and polarization, as a function of power for the $L= 1.19$~$\mu$m device.}
\end{center}
\end{figure} 

The temperatures $\bar{T}_\mx(P,\lambda,p)$ determined for a representative device are shown in Fig.~\ref{fig:TvsP}.  These temperatures are approximately linear with respect to the power applied to the nanotube over $\sim 700$~K. Excepting the data from the two longest wavelengths in the parallel polarization, at a given power the temperatures determined at different wavelengths fall within a range 60--160~K for all of the devices studied. Again, error analysis of, \emph{e.g.} Eq.~(\ref{eq:simple}) indicates that the uncertainty in a single-color temperature determination is proportional to the wavelength (and temperature), so long wavelengths (and high temperatures) are expected to give the least reliable temperature determinations.  However, the magnitude of the discrepancy at 1050~nm and 1100~nm in the parallel polarization is not understood.

\begin{figure}
\begin{center}
\includegraphics[width= \columnwidth]{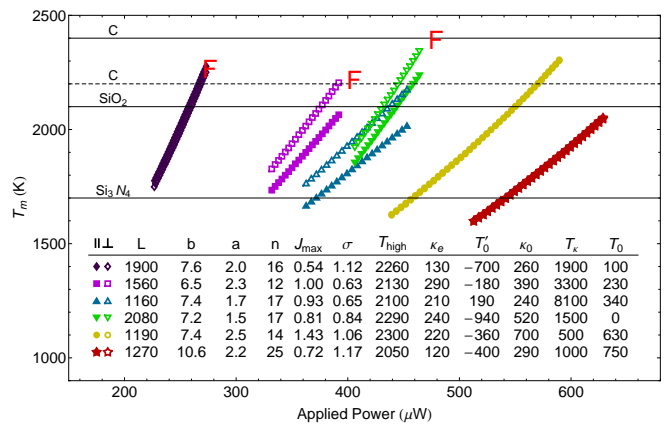}
\caption{\label{fig:zoo} (Color online) Plot of averaged temperatures $\bar{T}_\mx(P,p)$ versus power $P$ for 6 devices, as determined for both polarizations $p$ separately and assuming graphene's optical conductivity.  Temperatures corresponding to thickness loss rates of 1~nm/s for various materials are indicated with solid horizontal lines.  An additional dashed horizontal line indicates a loss rate of one graphene monolayer per minute.   The inset table shows for each device the final best estimates for the highest $T_\mx$ achieved, $T_\text{high}$, which is computed by adjusting the conductivity $\sigma$ (in units of graphene's optical conductivity) to a value such that $\bar{T}_\mx(P,||)$ and  $\bar{T}_\mx(P,\perp)$ are in agreement.  Other nanotube parameters are also given: the length $L$, the outer radius $b$, the inner radius $a$ (all in nm), the number of walls $n$, and the maximum current density $J_\text{max}$ (in $\mu$A/nm$^2$).  The $\kappa$'s (in $\text{W}/\text{m}\cdot\text{K}$) and other $T$'s (in K) result from fits to determine the thermal conductivity as described in the main text. Three devices failed while under observation at the powers and temperatures indicated by the labels ``F''.}
\end{center}
\end{figure}

To find $\bar{T}_\mx(P,p)$ we weight $\bar{T}_\mx(P,\lambda,p)$ according to an error $\propto \lambda$. Figure~\ref{fig:zoo} shows $\bar{T}_\mx(P,p)$ for 6 different devices with lengths ranging from $1160$~nm to $1900$~nm and wall numbers between 13 and 25. For each device the weighted standard deviation across wavelengths is $\lesssim 100$~K, which we take to be our uncertainty. For these devices the average difference between $\bar{T}_\mx(P,||)$ and $\bar{T}_\mx(P,\perp)$ is 70~K with no discrepancies greater than 160~K, also consistent with our claimed uncertainty.  

The degree of consistency between the temperature determinations for different polarizations, wavelengths, and positions on the CCD is not necessarily an indication of their accuracy, for it could reflect a systematic error. The dominant source of systematic uncertainty here is the emission efficiency $Q$, which we have calculated from reasonable but approximate first principles.  We assign the $Q$ calculation an uncertainty $\delta Q/Q \sim 0.5$, reflecting our limited \emph{a priori} knowledge of the nanotubes' optical conductivity.  Such an uncertainty implies $\delta T_\mx\sim (\lambda T_\mx^2/C_2)(\delta Q/Q)\lesssim 90$~K, consistent with the range we find experimentally.  However, because $Q_{||}$ and $Q_\perp$ happen to have such dissimilar dependences on the optical conductivity, the consistency of the determinations of $\bar{T}_\mx(P,||)$ and $\bar{T}_\mx(P,\perp)$ provides a powerful, independent validation of both the initial estimate of the optical conductivity and the claimed uncertainty.  

It is not necessary to assume an initial value for the optical conductivity $\sigma$ that appears in the Mie calculation of the efficiencies.  Treating $\sigma$ as a free parameter, we find that it can only vary in a narrow range and still accomodate the data in both polarizations. The table inset in Fig.~\ref{fig:zoo} shows how $\sigma$, in units of the graphene value $\sigma_g/\delta\simeq 1/(5590\,\mu\Omega\,\text{cm})$, must be adjusted to bring  $\bar{T}_\mx(P,||)$ and $\bar{T}_\mx(P,\perp)$ into agreement.  The adjustment is small in all cases: typically $\lesssim 15\%$, with a most discrepant value of 37\%.  Thus, barring breakdown of the conducting tube model of the nanotube filament, consistency requires that these nanotubes have optical conductivities in a relatively narrow range around $\pi \alpha/(Z_0 \delta)$, and that the efficiencies $Q$ and temperatures $T_\mx$ which logically follow have the claimed values to within the given uncertainties.

In addition to presenting an internally consistent picture, the derived temperatures are reasonable considering the observed physical modifications to the membrane and the nanotubes themselves. Graphite, silicon nitride, and silicon dioxide are observed to sublime in vacuum at high temperatures, with thickness loss rates of $\sim 1$~nm/s at $2400$~K, $1700$~K, and $2100$~K respectively. \cite{1996Rocabois,2002Haines,1990Hashimoto} (TEM energy-dispersive X-ray spectroscopy measurements indicate that the membranes contain some residual SiO$_2$ in addition to the Si$_3$N$_4$.) For graphite and silicon nitride these loss rates --- substantial for nanoscale devices --- occur many hundreds of kelvin below their melting temperatures.  At $2200$~K graphite evaporates at a rate corresponding to one graphene layer per minute, which is roughly the timescale of our longest exposures (2~min).  The three nanotubes labeled ``F'' in Fig.~\ref{fig:zoo} failed while under observation at temperatures of 2260, 2210, and 2390~K respectively. (In the latter two cases the temperatures represent extrapolations of $\sim 100$~K since the devices failed shortly after 0.1~V increases in the applied bias.)  All six devices showed substantial, thermally-induced holes (lengths $0.3$--$0.7~L$) in their silicon nitride membranes. Thus the decomposition temperatures seen in the nanolamp devices agree with those expected for such materials.

\begin{figure}
\begin{center}
\includegraphics[width= \columnwidth]{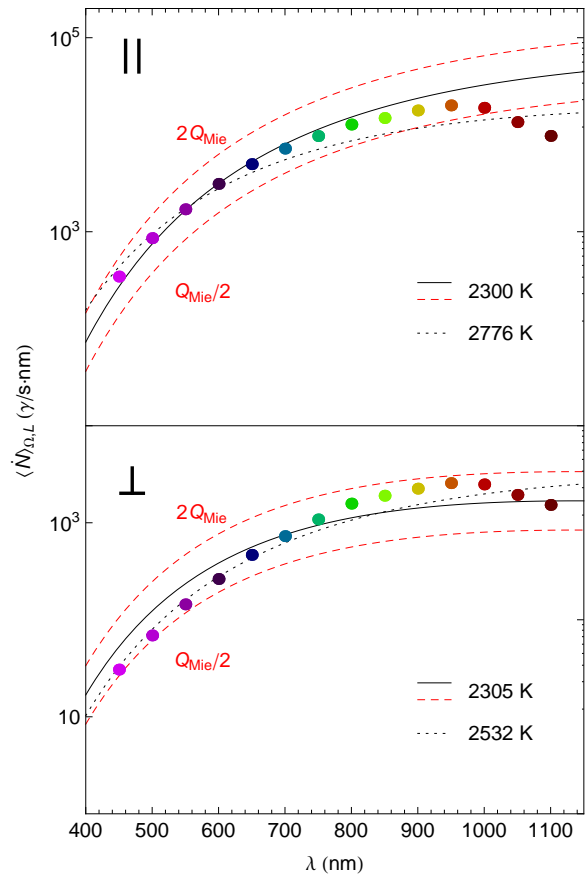}
\caption{\label{fig:emit}(Color online) Logarithmic plots of the average photon emission rate  per solid angle per unit length in both polarizations from the $L=1.19$~$\mu$m device.  The dots, colored as in Fig.~\ref{fig:TvsP}, are the maximum power data from Eq.~(\ref{eq:measured}). The solid curves give the single-color pyrometry expectation described by Eq.~(\ref{eq:expected}), which is based on the Mie calculation of the efficiency $Q$ and the temperatures shown in Fig~\ref{fig:zoo}.  The dashed curves show how the rates change if the $Q$'s are altered by a factor of 2.  The dotted curves are the best greybody fits using the multiwavelength method.}
\end{center}
\end{figure}

To compare single-color pyrometry with the more common multiwavelength pyrometry, we take the filament to be grey and vary both the efficiency $Q$ and $T_\mx$ to best fit the data at the 14 different wavelengths.  Here we sum over pixels and divide by the solid angle captured for easier comparison with a non-imaging, small NA system.  The measured number of photons per unit solid angle per unit length is then
\begin{equation}\label{eq:measured}
\langle \dot{N} \rangle_{L,\Omega} = \frac{\sum_i \sum_j \dot{S}(i,j)}{\Theta_f L \int_\text{NA} d\Omega}.
\end{equation}
In Fig.~\ref{fig:emit} we plot this data for a representative device, along with the single-color pyrometry expectation
\begin{equation}\label{eq:expected}
\langle \dot{N} \rangle_{L,\Omega} = \frac{2 b c \Delta \lambda}{\lambda^4} \frac{\int_\text{NA} Q_\text{abs} d\Omega}{\int_\text{NA} d\Omega} \frac{\int_{-L/2}^{L/2} e^{-C_2/\lambda T(x)} dx}{L},
\end{equation}
where we take $T_0=300$ and $T_\mx$ from the analysis used to produce Fig.~\ref{fig:zoo}. Here we include all of the geometric factors required to give the absolute signal levels; these factors are not required for the multiwavelength analysis, which more commonly has an ordinate labeled in arbitrary units. Two curves reflecting changes to $Q$ by a factor of 2 and one curve representing the multiwavelength analysis are also shown for each polarization.

As expected, the two-parameter multiwavelength fit is visibly superior to the single-color direct calculation, within a given polarization.  However, the agreement between the temperatures found for the two polarizations is markedly inferior.  For the six devices shown in Fig.~\ref{fig:zoo}, single color pyrometery gives $\bar{T}_\mx(P_\text{max},||)-\bar{T}_\mx(P_\text{max},\perp)$ of 70~K on average, with a standard deviation 70~K and a maximum discrepancy of 160~K.  For multiwavelength pyrometry the equivalent numbers are 190~K, 110~K, and 330~K, which is to say that the consistency between polarizations of multicolor pyrometry is worse by a factor $\sim 2$.  The multiwavelength results also give $T_\text{high} \sim 2700$~K, corresponding to a vapor pressure of carbon above graphite of $10^{-6}$ atm, or an erosion rate of 10~nm/s.\cite{2002Haines}  Thus, if a single temperature characterizes the radiation intensity for both polarizations and these MWCNTs are not markedly more refractory than the thermodynamic ground state of carbon, the single-color pyrometric method gives results that are both more internally consistent and more physically reasonable.  The larger temperatures \footnote{Since we are operating in the Wien limit, the high temperatures returned by the multiwavelength analysis imply that the emissivity model used (the greybody model) overestimates the intensity at long wavelengths relative to that at short wavelengths.  The observed deficits at the red end of the spectrum thus contribute to an overestimation of the temperature.  We find that reasonable, non-grey emissivity models with adjustable parameters return temperatures that, depending on the functional form chosen, range over more than 1500~K, which again emphasizes the advantage having a fixed emissivity model.} returned by the multiwavelength analysis also require that both $Q^{||}$ and $Q^\perp$ be smaller by an order of magnitude to give the observed signal levels. Within our simple model for the dielectric constant, such small efficiencies cannot be simultaneously arranged with any value of the optical conductivity.

With the final temperature determinations and the nanotubes' dimensions we  extract values for the MWCNTs' thermal conductivity $\kappa$ using two different models.  To compare with previous work, we fit the Fig.~\ref{fig:zoo} data to the function $T_\mx = T_0^\prime + m P$. The maximum temperature $T_\mx$ increases roughly linearly with the applied power $P$, consistent with a temperature-independent $\kappa$.  Approximating the nanotube resistivity to be also temperature independent gives the slope $m$ in terms of the thermal conductivity, $m = L/(8 \kappa_{\text{e}} \pi (b^2-a^2))$ (Refs.~\onlinecite{2001Collins,2005Chiu,2009Fan}).  We find $\kappa_{\text{e}} = 100-300~\text{W}/\text{m}\cdot\text{K}$ (see table inset in Fig.~\ref{fig:zoo}), consistent with previous experiments that have applied various thermometric assumptions to strongly-biased individual MWCNTs and have found values ranging from 50 to 600~W/m$\cdot$K \cite{2001Collins,2005Chiu,2007Begtrup1}.  The values $\kappa_{\text{e}}$ of the effective thermal conductivity correspond to an average over the range from 300~K to 2300~K, since $\kappa$ is not constant but rather decreasing above room temperature\cite{2007Begtrup1}.

The slight curvature in the curves $\bar{T}_\mx(P,p)$ shown in Fig.~\ref{fig:zoo} and the consistently low values for the zero-power temperature $T'_0$ are evidence of this temperature variation. We build a more complete model by adopting the functional form for the thermal conductivity $\kappa(T)=\kappa_0/(1+(T-T_0)/T_\kappa)$, which can describe both the constant $\kappa$ and Umklapp scattering-dominated $\kappa\propto 1/T$ cases with suitable parameter choices.  The steady-state heat equation\cite{2009Fan} is then non-linear,
\begin{equation}
 0=\frac{\partial}{\partial x} (\kappa(T)\, \pi(b^2-a^2) \frac{\partial}{\partial x} T)+I^2 \frac{\rho}{\pi(b^2-a^2)}
 \end{equation}
with the Gaussian, as opposed to parabolic, solution
\begin{equation}\label{eq:heatsolution}
T(x)=T_\kappa\left( \exp\left[\frac{P L(1-4 x^2/L^2)}{8\pi(b^2-a^2)\kappa_0 T_\kappa} \right] -1\right)+T_0.
\end{equation}
Here the boundary condition  $T(x=\pm L/2)=T_0$ has been enforced.  Near $x=0$ where the nanotube is brightest Eq.~\ref{eq:heatsolution} is approximately parabolic, so the assumption underlying Eq.~\ref{eq:counts} is not invalidated.

We fit the $\sigma$-adjusted data of Fig.~\ref{fig:zoo} to $T_\mx(P)$ as described by Eq.~\ref{eq:heatsolution} at $x=0$ with $\kappa_0$, $T_\kappa$, and $T_0$ as free parameters.  The results given in the table inset in Fig.~\ref{fig:zoo} show thermal conductivities that decrease with temperature for all of the devices, with a ratio $\kappa_0/\kappa(T_\mx=2000\,\text{K})\sim 2$.  Although this more sophisticated analysis still neglects effects such as the temperature variation of the resistivity $\rho$ and the power dissipation in the contacts, it gives a remarkably improved extrapolation to room temperature, returning $T_0 \simeq 340$~K for the temperature of the contacts on average.  Thus  the midpoint temperature $T_\mx$ is more accurately described as exponential, not linear, in the power $P$ over the wide range from 300 to 2300~K.\\

\section{Conclusion}
In conclusion, we have used single-color optical pyrometry to determine the temperatures attained by a half-dozen individual MWCNTs Joule-heated to incandescence.  This thermometry technique requires absolute calibration of the light collection apparatus and absolute knowledge of the thermal emission cross section of the MWCNTs. The physical cross sections are determined using TEM, and the emission efficiencies are calculated by solving Maxwell's equations for a conducting tube.  Since the MWCNTs have radii which are much smaller than a thermal wavelength, the emission efficiency model inherently assumes phase coherence across the radiating volume.  Comparing the temperature values determined across pixels, wavelengths, and polarizations shows that the model is internally consistent, and more consistent than one based on multiwavelength pyrometry. Furthermore, because the nanolamp emission efficiency increases in one polarization as it decreases in the other as a function of the optical conductivity, the nanolamp optical conductivity is completely determined by the degree of polarization observed in the data.  Experimentally we find that $\sigma=0.9\pm 0.2$ in units of idealized graphene's optical conductivity $\pi \alpha/(Z_0 \delta)$.  This conductivity fixes the emission efficiency, which in turn determines the temperatures.   The nanolamps' observed lifetimes and derived temperatures are consistent with a sublimation failure mode.  The MWCNTs have thermal conductivities ranging from  $400\pm 200~\text{W}/\text{m}\cdot\text{K}$ at room temperature to $200\pm 100~\text{W}/\text{m}\cdot\text{K}$ at 2000~K. These values, while extraordinarily high compared to most materials at elevated temperatures, are consistent with previous measurements on this material. Thus we find that optical pyrometry based on absolute measurements of visible incandescence gives a consistent picture of structurally-characterized radiators that are both large enough to have bulk electronic properties and small compared to a thermal wavelength.
\acknowledgments
This project is supported by NSF CAREER Grant No. 0748880.  All TEM work was performed at the Electron Imaging Center for NanoMachines at UCLA.
\bibliography{blackpaper}

\end{document}